\documentclass[10pt,preprint2]{aastex}

\def\Msol {\hbox{M$_{\odot}$}}

\def\kms {\hbox{${\rm km\, s}^{-1}$}}
\def\acs {\hbox{$^{\prime\prime}$}}

\def\obc {$\rm [OBC97] $}
\def\IRAM {IRAM 30m}
\input{epsf}
\slugcomment{To be submitted to the Astrophysical Journal Letters}
\shortauthors{O'Neil \& Schinnerer}
\shorttitle{Molecular Gas in MLSB Galaxies}
\begin{document}
\title{Properties of Molecular Gas in Massive Low Surface Brightness Galaxies, \\
Including New $^{12}$CO Observations of Three Malin 1 `Cousins'}
\author{K. O'Neil}
\affil{NRAO, P.O. Box 2, Green Bank, WV 24944 USA}
\email{koneil@nrao.edu}

\and
\author{E. Schinnerer\altaffilmark{1}}
\affil{NRAO, P.O. Box 0, Socorro, NM 87801 USA}
\altaffiltext{1}{Jansky Postdoctoral Fellow at the National Radio Astronomy Observatory}
\email{eschinne@nrao.edu}

\begin{abstract}
To date, the only low surface brightness (LSB) galaxies which have been detected in CO
are the Massive LSB (MLSB) galaxies.  
In 2003, O'Neil, Schinnerer, \& Hofner hypothesized
that is the prominent bulge component in MLSB galaxies, not present
in less massive low surface brightness galaxies, which gives rise to
the detectable quantities of CO gas.  To test this hypothesis, we have used
the IRAM 30m telescope to obtain three new, deep CO J(1$-$0) and J(2$-$1)
observations of MLSB galaxies.  Two of the three galaxies
observed were detected in CO -- one in the J(1$-$0) line and the other in
both the J(1$-$0) and J(2$-$1) lines, bringing the total number of MLSB galaxies
with CO detections to 5, out of a total of 9 MLSB galaxies observed at CO to date.  
The third object had no detection to 2 mK at CO J(1$-$0).
Comparing all MLSB galaxy CO results with surveys
of high surface brightness galaxies, we find 
the MLSB galaxies' M$_{H_2}$ and M$_{H_2}$/M$_{HI}$ values
fall within the ranges typically 
found for high surface brightness objects, albeit at the low end of the distribution,
with the two MLSB galaxies detected at CO in this survey having the highest M$_{H_2}$/M$_{HI}$ values
yet measured for any LSB system, by factors of 2--3.
\end{abstract}

\section{Introduction}

Despite  more than a decade of study, the star formation processes within
low surface brightness (LSB) galaxies, those objects with central surface brightness at least
1 magnitude fainter than the night sky (uncorrected for inclination), remain enigmatic.
The general properties of LSB galaxies -- blue colors, high gas
mass-to-luminosity ratios, and low metallicities -- lead to the conclusion
that LSB systems are under-evolved compared to their high surface brightness
(HSB) counterparts.   When combined with the low gas density
(typically $\rm \rho_{HI} \leq 10^{21}\;cm^{-2}$) and low
baryonic-to-dark matter content typical of LSB systems, the question
can be raised not of why LSB galaxies are under-evolved, but instead
of how LSB systems form stars at all (O'Neil, Schinnerer, \& Hofner 2003,
and references therein).

One of the primary methods for studying the star formation rate and
efficiency in galaxies is through study of the galaxies' interstellar medium (ISM), and
one mechanism for studying a galaxy's ISM is
through observing its CO content.  Until recently, all attempts at detecting
CO in LSB systems had been unsuccessful \citep{braine00,deblok98,schombert90},
leading to speculation as to why LSB galaxies appear to lack molecular
gas.  However, in the past three years the first detections of CO in LSB galaxies 
have been made, giving the first look into
the ISM of LSB systems \citep{oneil03,oneil03b,oneil00}. 

While the three detections to date of CO gas in LSB galaxies are of considerable importance,
it is clear further study is paramount to understand LSB systems -- knowing the CO content 
of only three LSB galaxies does not provide enough information to understand the molecular gas content of
LSB systems as a whole.   To aid in our understanding, we undertook CO observations
of three additional LSB galaxies, with properties similar in most respects to the properties
of the LSB galaxies with detected CO.  The results of those observations are described herein.

\begin{deluxetable}{lccccccccc}
\tabletypesize{\scriptsize}
\tablecolumns{10}
\footnotesize
\tablewidth{0pt}
\tablecaption{Known Properties of MLSB Galaxies Observed in CO \label{tab:props}}
\tablehead{
\colhead{Name}  & \colhead{Type} &  \colhead{$\rm \mathbf \mu_B (0) $} &
\colhead{M$_{B}$} & \colhead{D$_{25}$}&
\colhead{\it  i} & \colhead{log(${M_{HI}}\over{M_\odot}$)} &
\colhead{$\rm v_{HEL}^{HI}$} & \colhead{$\rm w_{20, obs}^{HI}$} & \colhead{Neighbors$^a$}\\
\colhead{} & \colhead{}  & \colhead{[mag arcsec$^{-2}$]}& \colhead{[mag]}& \colhead{[kpc]} &
\colhead{[$^\circ$]} & \colhead{}  & \colhead{[km s$^{-1}$]}  & \colhead{[km s$^{-1}$]} &
}
\startdata
\cutinhead{Observations from this paper}
UGC 04144     & Sc     & 24.4$^{1,b}$ & -20.0     & 42 & 83 & 9.9$^{1}$   & 9795$^{1}$  & 494$^{1}$ & 1\\
UGC 05440     & Sd     & 25.7$^{1,b}$ & -20.5     & 96 & 65 & 10.8$^{1}$  & 18932$^{1}$ & 531$^{1}$ & 0\\
UGC 06124     & S      & 26.0$^{1,b}$ & -19.9     & 81 & 82 & 10.3$^{1}$  & 13970$^{1}$ & 583$^{1}$ & 0\\
\cutinhead{Previous Detections}
UGC 01922     &S?      &\nodata$^c$ &-19.8        & 59 &38  & 10.33$^{1}$ &10894$^{1}$  & 1120$^{1,d}$& 2\\
UGC 12289     &Sd      &23.3        &-19.7        & 57 &22  & 10.13$^{1}$ &10160$^{1}$  & 488$^{1}$ & 2\\
\obc\ P06-1   &Sd      &23.2        &-18.6        & 29 &70  & 9.87        &10882        & 458       & 1\\
\cutinhead{Previous Non-detections}
UGC 06968     &Sc      &\nodata$^c$ &-21.1        & 48 &71  & 10.30       & 8232        & 574       & 3\\
LSBC F582-2   &Sbc     &\nodata$^c$ &\nodata      &41  &66  & 9.99        & 7043        & 310       & 1\\
Malin 1       &S       &26.4        &-21.4        &240 &20  & 10.6        &24733        & 710       & 0\\
\enddata
\tablecomments{
Unless otherwise noted, properties are from O'Neil, Schinnerer, \& Hofner (2003) and references therein.\\
$^a$The number of galaxies within a 750 kpc/2,000 \kms\ radius, as found with NED.
$^b$A measured $\mu_B(0)$ for this galaxy is not available.  The central surface brightness is
defined as $\rm \mathbf \langle \mu_B \rangle = m_{pg}+5\log{(D)}+8.89-0.26$,  where $m_{pg}$ is
the photographic magnitude from the UGC, D is the diameter in arcmin, 8.89 is the conversion from
arcmin to arcsec, and 0.26 is an average conversion from m$_{pg}$ to m$_B$ \citep{bothun85}.\\
$^c$The classification of this galaxy as an LSB galaxy is from \citet{schombert98}.\\
$^d$The velocity width for UGC 01922 uncorrected for inclination is 690 \kms \citep{giovanelli85}.
%$^c$The reported central surface brightness is through the V (not B) band \citep{pildis97}.
}
\tablerefs{
$^1$\citet{oneil04};
%$^2$O'Neil, Schinnerer, \& Hofner (2003) and references therein;
%$^3$\citet{oneil00};
%$^1$\citet{schombert98};
%$^2$\citet{nilson73};
%$^3$\citet{giovanelli85};
%$^5$\citet{oneil00b};
%$^6$\citet{oneil97a};
%$^7$\citet{oneil97b};
%$^{8}$\citet{boselli94};
%$^{9}$\citet{gavazzi87};
%$^{10}$\citet{schombert88};
%$^{11}$\citet{garnier96};
%$^{12}$\citet{eder00};
%$^{13}$\citet{schombert90};
%$^{14}$\citet{deblok96};
%$^{15}$\citet{schombert92};
%$^{16}$\citet{mcgaugh94};
%$^{17}$\citet{huchtmeier00};
%$^{18}$\citet{deblok95};
%$^{19}$\citet{deblok97};
%$^{20}$\citet{impey89}
}
\end{deluxetable}

\begin{deluxetable}{lccccc}
%\scriptsize
\tablecolumns{6}
%\footnotesize
\tablewidth{0pt}
\tablecaption{Parameters for IRAM Observations \label{tab:galobs} \label{tab:obs}}
\tablehead{
\colhead{Name} & \colhead{RA}& \colhead{DEC}& 
\colhead{$\sigma_{rms}^{1-0}$\dag}& \colhead{$\sigma_{rms}^{2-1}$\dag} \\
& \colhead{[J2000]}& \colhead{[J2000]}& \colhead{[mK]}&
\colhead{[mK]}
}
\startdata
UGC 04144 & 07:59:27.3 & 07:26:30.0 & 3.4 & 2.0 \\
UGC 05440 & 10:05:35.9 & 04:16:45.0 & 1.7 & 3.4 \\
UGC 06124 & 11:03:39.5 & 31:51:30.0 & 2.0 & 2.6 \\
\enddata
\tablecomments{\dag R.m.s. estimates are from the smoothed data (26.9 \kms\ resolution).}
\end{deluxetable}

\section{Observations}

As mentioned above, the three sources observed for this project have properties similar to
those of the three LSB galaxies which previously have been detected in
CO.  That is, all three galaxies observed fall into the category 
of Massive LSB (MLSB) galaxies, or `Malin 1' cousins (named after the largest and most 
famous of the MLSB galaxies), with M$_{HI} \ge 10^{10}$ \Msol,
W$_{20}^{corr} \ge$ 450 \kms, M$_B \le -$18.5, and D$_{25} \ge$50 kpc.
A complete description of the previously known properties of all MLSB galaxies
observed at CO, including those observed for this paper are given in 
Table~\ref{tab:props}. 

The CO J(1--0) and J(2--1) rotational transitions of the galaxies were observed using
the \IRAM\ telescope in the period from 7-8 March, 2003.
Table~\ref{tab:galobs} lists the adopted positions (determined using the
digitized Palomar sky survey plates and accurate to 2-3\acs) and
heliocentric velocities \citep{oneil04} for our target sources.
The beams (22\acs\ at 110 GHz) were centered on the nucleus of each galaxy.
This resulted in a coverage of the inner 14, 27, and 24 kpc for UGC 04144, UGC 05440,
and UGC 06124, respectively.
Pointing  and focus were checked every hour
and pointing was found to be within the telescope limits (better than $2^{\prime\prime}$).
For each source both transitions were observed
simultaneously with two receivers.
Both back ends were set using with 510 MHz bandwidth and 1.25 MHz resolution,
resulting in an unsmoothed resolutions of 3.36 and 1.68 \kms\ for the 3mm 
and 1mm observations, respectively.
For data reduction, the lines were smoothed
to 26.9 \kms\ resolution.
Each target was observed on-source for a total of 66 minutes.

All observations used the wobbling secondary with
the maximal beam throw of $240^{\prime\prime}$.
The image side band rejection ratios were measured
to be $>30$dB for the $3\,$mm SIS receivers and
$>12$dB for the $1.3\,$mm SIS receivers.  The data were
calibrated using the standard chopper wheel technique \citep{kut81}
and are reported in main beam brightness temperature T$_{MB}$.
Typical system temperatures during the observations
were 170--190K and 350--450K in the $3\,$mm and $1.3\,$mm band, respectively.
All data reduction was done using CLASS -- the Continuum and Line Analysis
Single-dish Software developed by the Observatoire de Grenoble and IRAM \citep{buisson02}.

\begin{figure}[h]
\plotone{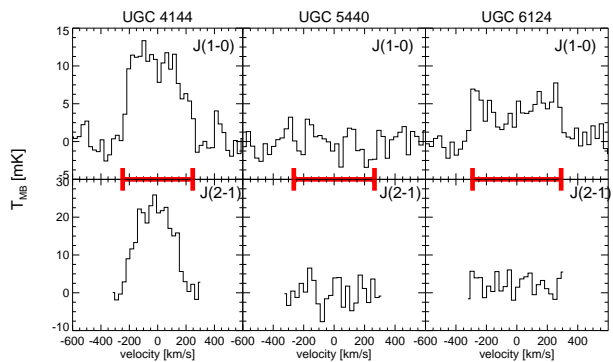}
\caption{IRAM 30m CO J(1$-$0) and J(2$-$1) spectra for UGC~04144, UGC~05440, and UGC~06124.
The data have been smoothed to a resolution of 26.9 \kms.
The horizontal bars indicate the extent of the observed (uncorrected) \ion{H}{1}
velocity widths (at 20\% the peak HI intensity) as given by \citet{oneil04}. \label{fig:u4144}}
\end{figure}

\begin{figure}
%\plottwo{f2a.eps}{f2b.eps}
%\plottwo{MBH2.eps}{MBH2HI.eps}
%\plottwo{f2c.eps}{f2d.eps}
%\plottwo{v20cH2.eps}{v20cH2HI.eps}
\plotone{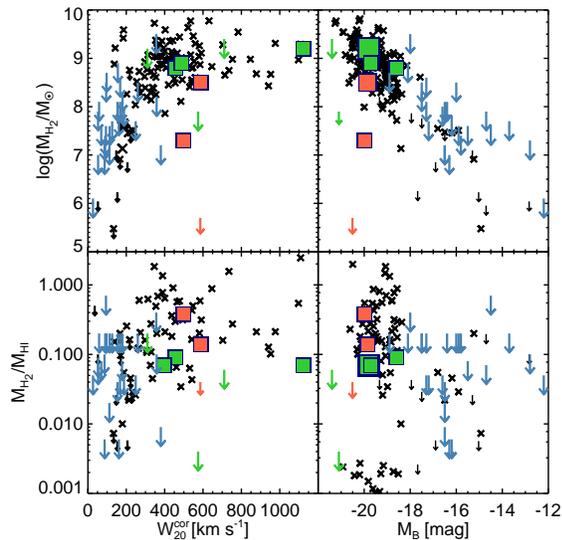}
\caption{Inclination corrected \ion{H}{1} velocity widths versus
H$_2$ mass (top left) and the H$_2$-to-\ion{H}{1} mass
ratio (bottom left).  At right is the absolute B magnitude versus H$_2$ mass 
(top right) and the H$_2$-to-\ion{H}{1} mass
ratio (bottom right).  The red symbols are LSB galaxies from this survey, the
green symbols are MLSB measurements from previous surveys \citep{oneil03, oneil00}, 
the blue are previous LSB (lower mass) measurements \citep{oneil00,braine00, deblok98, schombert90}, 
and the black symbols are taken from various studies of the CO content in
HSB spiral galaxies \citep{casoli96, boselli96, tacconi87} and from the \citep{matthews01}
study of extremely late-type, edge-on spiral galaxies.
An arrow indicates only an upper limit was found.  Note that only 5 MLSB galaxies
are shown in the plots on the right, as the absolute magnitude of LSBC F582-2 is
not known.
\label{fig:MBH2}}
\end{figure}

\section{Results}

Two of the three galaxies observed, UGC 04144 and UGC 06124, were detected in CO
while the third galaxy, UGC 05440, was not detected, with an upper limit of
$f_{CO(1-0)} <$ 0.82 K \kms and M$_{H_2} < 10^{9.2}$\Msol (Figure~\ref{fig:u4144}).
Due to its distance 
(v$_{HEL}$=18932 \kms), the upper M$_{H_2}$ limit placed on UGC 05440 was too high for any significance.
However, the limit placed on M$_{H_2}$/M$_{HI}$ is low for a galaxy of its luminosity.
Table~\ref{tab:CO} lists the CO properties of all three objects, as determined by our observations.  
For comparison with  previous studies (O'Neil, Schinnerer, \& Hofner 2003, and references therein)
%\citep{oneil03,burkholder01,oneil00,bergvall99,
%deblok98,roennback95, mcgaugh94,vdhulst93,davies90},
we used a standard CO $\leftrightarrow$ H$_2$
conversion factor (X) of $\rm N(H_2)/\int{T(CO)dv}\;=\;3.6\;\times\;10^{20}
\;cm^{-2}/(K\;km\;s^{-1})$ adopted from \citet{sanders86}.  As discussed in \citet{oneil00},
this assumption does not include dependence based on the structure of the ISM, metallicity,
etc.  A discussion of any errors which may arise due to this assumption can be found in 
\citet{oneil03}.

Of the three galaxies observed, UGC 04144 is the most nearby, making it unsurprising that
UGC 04144's J(1$-$0) flux is considerably higher than that found for the other galaxies.
However, it is notable that both UGC 04144 and UGC 06124 have M$_{H_2}$/M$_{HI}$ values
considerably higher than is found for the other MLSB galaxies with CO detections.
Before these observations, the average value of M$_{H_2}$/M$_{HI}$ for the MLSB galaxies
was 0.08.  In contrast, UGC 06124 has M$_{H_2}$/M$_{HI}$ = 0.14 and UGC 04144 has
M$_{H_2}$/M$_{HI}$ = 0.38.  Comparing these galaxies with the other MLSB galaxies
observed, both those with and without CO detections, shows very little differences in their properties.
All galaxies observed have similar morphologies (Sc/Sd), colors (from U through K -- 
NED\footnote{NED is the NASA Extragalactic Database, available online at http://nedwww.ipac.caltech.edu/}), 
\ion{H}{1} masses, 
and total (dynamic) masses.  Two of the MLSB galaxies have both 1.4 GHz continuum and IRAS detections
-- UGC 04144 (f$_{1.4GHz}$=4.7 mJy, f$_{60\mu m}$=0.40 Jy) and UGC 01922
(f$_{1.4GHz}$=38.5 mJy), f$_{60\mu m}$=0.33 Jy \citep{condon98, beichman88, moshir92}, but UGC 01922 has 
M$_{H_2}$/M$_{HI}$ of only 0.07.  Similarly, the number of neighboring galaxies does not seem
to alter the quantity of molecular in within these galaxies.  UGC 04144 has one nearby neighbor
(NGC 02499, at a distance of 350 kpc and $\Delta v$=185 \kms), while UGC 06124 has no galaxies within
a 750 kpc/2,000 \kms\ radius (Table~\ref{tab:props}).

The single quantity which does appear to distinguish UGC 04144
and UGC 06124 from the other MLSB galaxies observed is their high inclination 
({\it i}=83$^\circ$ and 82$^\circ$ for UGC 04144 and UGC 06124, respectively, versus
{\it i}=22$^\circ$, 38$^\circ$, and 70$^\circ$ for the other three galaxies with CO
detections).  However, the error for the inclination measurements is 5--10$^\circ$,
making the inclination of UGC 04144/UGC 06124 comparable to that of [OBC97] P06-1 ({\it i}=70$^\circ$),
UGC 06968 ({\it i}=71$^\circ$).  As a result, while it is possible the high inclination angle 
has contributed to the higher M$_{H_2}$/M$_{HI}$ values seen for UGC 04144 and UGC 06124, it is unlikely
this is the only explanation.  Follow-up \ion{H}{1} and CO imaging should help resolve this question.

\section{Conclusions -- Molecular Gas in LSB Galaxies}

With the results in this paper we have added two more measurements of molecular
gas in LSB galaxies, bringing the total number of detections up to five,
out of a total of nine MLSB (and 37 LSB galaxies of any type).
Figure ~\ref{fig:MBH2} compares the findings in this paper
with all other LSB galaxy CO studies and with a sample of measurements from a variety of other galaxy
studies.  These include `standard' HSB disk galaxy studies
\citep{boselli96,casoli96}, dwarf galaxy studies \citep{tacconi87},
and a study of extreme late-type spiral galaxies \citep{matthews01}.
In all cases a conversion factor of $\rm N(H_2)/\int{T(CO)dv}\;=
\;3.6\;\times\;10^{20}\;cm^{-2}/(K\;km\;s^{-1})$ was used to
allow ready comparison between the results.

As can be seen in Figure~\ref{fig:MBH2}, both the detected CO and upper limits placed on the
non-detections (and by inference H$_2$ detections and upper limits) for LSB galaxies fall within
the ranges typically found for high surface brightness objects.  (The one exception to
this, UGC 06968, is described in detail in O'Neil, Schinnerer, \& Hofner 2003.)  Using the data from 
\citet{casoli96}, \citet{boselli96}, and \citet{tacconi87}
gives $\langle M_{H_2}/M_{HI} \rangle$=0.51$\pm$0.78
for all HSB galaxies and $\langle M_{H_2}/M_{HI} \rangle$=0.53$\pm$0.80 for HSB galaxies
with M$_B <-18.5$.  These numbers, though, are skewed due to the presence of a few galaxies
with $M_{H_2}/M_{HI} > 1$.  Looking instead at the median value for the HSB galaxies with M$_B <-18.5$
is $\langle M_{H_2}/M_{HI} \rangle_{median}$=0.27.  The 
MLSB galaxies with CO detections have $M_{H_2}/M_{HI}$=0.07 -- 0.5, with 
$\langle M_{H_2}/M_{HI} \rangle_{median}$=0.09, within the range of the values for the HSB
galaxies, albeit a bit lower.  As no correction has been made to the MLSB galaxy data to account
for surveying only the central 10-25 kpc of each galaxy for CO, the fact that the MLSB galaxies'
M$_{H_2}$/M$_{HI}$ appears to be somewhat lower than that found for the HSB galaxies cannot
be considered significant.

It is also clear from Figure~\ref{fig:MBH2} that the only LSB galaxies
which have been detected at CO are the massive LSB galaxies.
Unlike their less massive counterparts which often have little or no central
concentration of matter, the higher gravitational potential at the center of
MLSB galaxies typically results in a dense central bulge.  
\citet{oneil03} speculate 
that it is within this high density region that the star formation history of massive
LSB galaxies most readily mimics that of HSB galaxies, resulting in an overall higher
star formation rate, and producing the molecular gas detected.  This speculation
is given considerable more weight with our most recent observations.  Previously, detection
of molecular gas in LSB galaxies seemed like an impossible task, with the first 
CO detection occurring only after 10 years of searching.  Yet in this paper we described
observing only three LSB galaxies and detecting CO in two of the three -- a 67\% detection
rate. As all three sources were chosen using the criteria described in \citet{oneil03} -- LSB galaxies
with high dynamical masses, M$_B < -19$, and large central bulges -- it would appear
O'Neil et al.'s speculation has merit.  The high detection rate shows we 
are now able to reliably find CO gas within the central region of MLSB galaxies.  

\acknowledgements{Based on observations carried out with the IRAM 30m telescope. 
  IRAM is supported by INSU/CNRS (France), MPG (Germany) and IGN 
(Spain).   This research has made use of the NASA/IPAC Extragalactic Database (NED) which is operated by the Jet Propulsion Laboratory, California Institute of Technology, under contract with the National Aeronautics and Space Administration.}

\begin{deluxetable}{lcccccc}
\tabletypesize{\scriptsize}
\tablecolumns{7}
%\scriptsize
\tablewidth{0pt}
\tablecaption{Properties of All MLSB Galaxies Observed in CO\label{tab:CO}}
\tablehead{
\colhead{Galaxy} & \colhead{Line}&
\colhead{$\rm \int{T_{MB}dv}\dagger$}  &\colhead{$\rm v_{HEL}$}&
\colhead{Width} & \colhead{$\rm log(M_{H_2}/M_\odot$)$\ddagger$}& \colhead{$\rm M_{H_2}/M_{HI}$}\\
& & \colhead{[K km s$^{-1}$]}& \colhead{[km s$^{-1}$]} & \colhead{[km s$^{-1}$]} &&
}
\startdata
\cutinhead{Observations from this paper}
UGC 04144 & 1$-$0 & 3.56    & 9797    & 430     & 7.3     & 0.38    \\
UGC 04144 & 2$-$1 & 7.17    & 9763    & 429     & 9.0     & 0.20    \\
UGC 05440 & 1$-$0 & $<$0.50 & \nodata & \nodata & $<$5.7  & \nodata \\
UGC 05440 & 2$-$1 & $<$0.59 & \nodata & \nodata & $<$5.6  & \nodata \\
UGC 06124 & 1$-$0 & 1.72    & 13940   & 517     & 8.5     & 0.14    \\
UGC 06124 & 2$-$1 & $<$0.78 & \nodata & \nodata & $<$7.9  & \nodata \\
\cutinhead{Previous Detections}
UGC 01922$^1$    & 1$-$0& 1.38& 10795& 404& 9.2& 0.07\\
UGC 01922$^1$    & 2$-$1& 2.96& 10802& 403& 8.9& 0.04\\
UGC 12289$^1$    & 1$-$0& 1.16& 10162& 200& 9.0& 0.07\\
UGC 12289$^1$    & 2$-$1& 0.69& 10185& 201& 8.2& 0.01\\
\obc\ P06-1$^2$  & 1$-$0&0.95& 10904& 302& 8.8& 0.09\\
\obc\ P06-1$^2$  & 2$-$1&1.14& 10903& 216& 8.3& 0.03\\
\cutinhead{Previous Non-detections}
%\obc\ P07-1$^1$  & 1$-$0&$<$0.13&\nodata&\nodata&  $<$7.0& $<$0.006\\
%\obc\ P07-1$^1$  & 2$-$1&$<$0.032&\nodata&\nodata& $<$5.7& $<$0.003\\
%\obc\ N09-2$^1$  & 1$-$0&$<$0.18&\nodata&\nodata& $<$7.8& $<$0.006\\
%\obc\ N09-2$^1$  & 2$-$1&$<$0.33&\nodata&\nodata& $<$7.5& $<$0.003\\
%\obc\ P05-6$^1$  & 1$-$0&$<$0.22&\nodata&\nodata& $<$7.2& $<$0.009\\
%\obc\ P05-6$^1$  & 2$-$1&$<$0.39&\nodata&\nodata& $<$6.9& $<$0.005\\
%\obc\ C04-1$^1$  & 1$-$0&$<$0.37& \nodata&\nodata& $<$8.1& $<$0.02\\
%\obc\ C04-1$^1$  & 2$-$1&$<$0.88&\nodata&\nodata& $<$7.9& $<$0.01\\
UGC 06968$^1$    & 1$-$0&$<$0.21&\nodata&\nodata& $<$7.9& $<$0.004\\
UGC 06968$^1$    & 2$-$1&$<$0.58& \nodata&\nodata& $<$7.8& $<$0.003\\
LSBC F582-2$^2$  & 1$-$0&$<$0.54& \nodata&\nodata& $<$9.2& $<$0.2\\
Malin 1$^2$      & 1$-$0&$<$0.15& \nodata&\nodata& $<$9.4& $<$0.06\\
Malin 1$^2$      & 2$-$1&$<$0.35& \nodata&\nodata& $<$8.7& $<$0.01\\
\enddata
\tablecomments{$\dagger$Non-detection limits are
$I_{CO}\:<\:3{T_{MB}}{v^{20}_{HI}\over{\sqrt{N}}}$ (N = the number of channels,
T$_{MB}$ is the 1$\sigma$ rms main beam temperature).\\
$\ddagger$As described in Section 3, conversion to M$_{H_2}$ was done using
$\rm N(H_2)/\int T(CO)dv=3.6\times 10^{20}\;cm^{-2}/(K\;km\;s^{-1})$}
\tablerefs{$^1$\citet{oneil03}; $^2$\citet{oneil00}; %$^3$\citet{schombert90};
%$^4$\citet{deblok98}; $^5$\citet{braine00}.
}
\end{deluxetable}

%\onecolumn
%\begin{figure}
%\centerline{
%\epsfxsize=2.2in
%\epsffile{f3a.eps}
%\epsfxsize=2.2in
%\epsffile{f3b.eps}
%\epsfxsize=2.2in
%\epsffile{f3c.eps}}
%\caption{Images from the POSS II survey of UGC 04144, UGC 05440, and UGC 06124.
%The images are 3$^\prime$ across, and the black contour lines lie at approximately
%95\% of each galaxy's peak level.  Note that UGC 04144 has a foreground star, marked
%by the white circle. \label{fig:gals}}
%\end{figure}
%The black contour lise at $\sim$95\% of the peak level of the galaxy; 3' per side
%\onecolumn
%\begin{figure}
%\plottwo{f3a.eps}{f3b.eps}
%%\plottwo{v20cH2.eps}{v20cH2HI.eps}
%\caption{Inclination corrected \ion{H}{1} velocity widths versus
%H$_2$ mass (left) and the H$_2$-to-\ion{H}{1} mass
%ratio (right).  The red symbols are LSB galaxies from this survey, the
%green symbols are LSB galaxy measurements from previous surveys \citep{oneil03, oneil00,
%braine00, deblok98, schom90}, the blue symbols are from the \citet{matthews01}
%study of CO in extreme late-type spiral galaxies, and
%the black symbols are taken from various studies of the CO content in
%HSB spiral galaxies \citep{casoli96, boselli96, tacconi87}.
%An arrow indicates only an upper limit was found.
%\label{fig:W20}}
%\end{figure}
\twocolumn

\end{document}